\documentclass[rnote]{aa}
\usepackage{hyperref}        
\hypersetup{
  colorlinks   = true, 
  urlcolor     = blue, 
  linkcolor    = black, 
  citecolor   = black 
}
\usepackage{fixltx2e,longtable,ifpdf,amsmath,amssymb,natbib}
\ifpdf
  \usepackage{aeguill}
  \usepackage[pdftex]{graphicx,color}
\else
  \usepackage[dvips]{graphicx,color}
\fi

\usepackage{etoolbox}

\begin{document}

\title{Emission-line stars in the LMC:\\  the Armagh survey, and a
  metacatalogue}

\author{Ian D. Howarth}
\institute{Dept. of Physics \& Astronomy, UCL, Gower St., London WC1E
  6BT, UK\\
\email{idh@star.ucl.ac.uk}}

\date{Submitted }

 
  \abstract
    {}
    {Accurate astrometry is required to reliably cross-match
      20th-century photographic catalogues against 21st-century
      digital surveys.  The present work provides modern-era
      identifications and astrometry for the 801 emission-line objects
      ``of stellar appearance'' in the Armagh survey (the largest
      of its nature to date).}
    {Targets have been individually identified in digital images using
      the Armagh Atlas and, in most cases, unambiguously matched to
      entries in the UCAC astro\-metric catalogues.}
    {Astrometry with sub-arcsecond precision is now available for all the major
      photographic spectroscopic surveys of the LMC.  The
      results are used to compile an annotated metacatalogue of 1675
      individual, spectroscopically identified candidate H$\alpha$-emission stars,
      including detailed cross-matching between catalogues, and resolving many (though not
      all) identification ambiguities in individual primary sources.}
   {}


   \titlerunning{Emission-line stars in the LMC}
   \authorrunning{Ian D. Howarth}
   \maketitle
%

\section{Introduction}

The Large Magellanic Cloud (LMC) provides an unsurpassed laboratory
for the investigation of massive stars; its low reddening and relative
proximity facilitate both large-scale surveys and detailed studies of
individual objects of interest.  As a consequence, a number of
extensive photo\-graphic searches for luminous LMC stars with
H$\alpha$ emission were conducted as technological advances
allowed, first by \citeauthor{Henize56} (\citeyear{Henize56}; 172
`LH$\alpha$120-S' stars), and subsequently at Armagh
(\citealt{Lindsay63b}, \citealt{Andrews64}; 801 `L63' and `AL' stars)
and by \citeauthor{Bohannan74} (\citeyear{Bohannan74}; 625 `BE74'
stars), alongside more general surveys, such as Sanduleak's
\citep{Sanduleak70}.  At the time of writing, the Armagh and BE74
surveys still provide the two largest lists of spectroscopically
identified candidate LMC emission-line stars available.

\begin{table*}
\centering
\caption{Metacatalogue composition.}  
\begin{tabular}{l c c c c c c c c c}
\hline
&
\multicolumn{1}{c}{LH$\alpha$}&
\multicolumn{1}{c}{L63}&
\multicolumn{1}{c}{AL}&
\multicolumn{1}{c}{BE74}&
\multicolumn{1}{c}{BAT}&
\multicolumn{1}{c}{RPs}&
\quad&
\multicolumn{1}{c}{First}&
\multicolumn{1}{c}{Unique}\\
\hline
LH$\alpha$ 120-S&             \bf{172}& $\phantom{3}$53&   $\phantom{4}$72&               136&  $\phantom{1}$16&  $\phantom{5}$23&&               172&  $\phantom{3}$14\\
L63      &                &             \bf{358}&   $\phantom{33}$3&               137&  $\phantom{1}$19&  $\phantom{3}$45&&               305& 191\\
AL       &                &                &               \bf{446}&               172&  $\phantom{1}$31&  $\phantom{5}$34&&               372& 241\\
BE74     &                &                &                  &               \bf{624}&  $\phantom{1}$53&  $\phantom{5}$74&&               284& 242\\
BAT99    &                &                &                  &   $\phantom{4}$\emph{55}&              \bf{134}&  $\phantom{33}$0&&   $\phantom{3}$66&  $\phantom{3}$66\\
RPs      & $\phantom{1}$\emph{25}&                &                  &                  &                 &              \bf{576}&&               467& 467\\
\hline 

\end{tabular}
\label{tab:xmatch}
\begin{minipage}{1.0\textwidth}
$\qquad$
\newline
\tiny
{\bfseries Notes.}    
    LH$\alpha$~120-S = \citet{Henize56};
    L63 = \citet{Lindsay63b};
    AL = \citet{Andrews64};
    BE74 = \citet{Bohannan74};
    \mbox{BAT99 = \citet{BAT99};}
    RPs = \citet{Reid12}.\vskip 1mm

Entries show the number of stars
  common to each pair of primary catalogues (e.g., there are 172 entries in
  the LH$\alpha$~120-S listing, of which 53 are also in the L63 catalogue).
Twenty-five RPs catalogue entries have LH$\alpha$~120-S matches, but only 23
LH$\alpha$~120-S entries have RPs entries, because
\mbox{LH$\alpha$~120-S~24} corresponds to three separate RPs objects; similarly,
BE74-383 (HD~269828) matches three separate BAT99 entries.
  The last two columns give the number of `discovery' entries (e.g.,
  of the 446 entries in the AL catalogue, 372 had not been
  previously identified by \citealt{Henize56} or by \citealt{Lindsay63b}),
  and 
  the number of entries not appearing in
  any of the other tabulated catalogues.
The RPs study is targeted
on fainter objects than the 20th-century surveys, giving rise to
a large number of new targets, with relatively few recoveries of
known, brighter stars.   \vskip 1mm

  The BE74 total of 624 is one fewer than the number of their catalogue entries
  because \mbox{BE74 394} is not marked on their charts, and the
  counterpart is not 
  reliably identifiable.
The RPs total of 576 stars is three fewer than their asserted 579 objects
because their Table~A1 has only
  578 entries, including one duplicated listing (RPs1822) and one
  duplicated target (RPs447=RPs1014).
The sum of `First' discoveries (1666) is nine
  fewer than the total number of metacatalogue entries (1675),  because 
(i) LH$\alpha$~120-S~24 and BE74-383 are two `firsts' that translate to six objects,
and (ii) the metacatalogue
  includes five WR stars not included in the listed primary
  catalogues.
\end{minipage}
\end{table*}

The utility of all these catalogues has been limited by co-ordinates
that were originally given to only $\sim$arcminute precision -- inadequate
for reliable identifications based on position alone.  While
accompanying, relatively small-scale finder charts normally provide
a more secure route to identification, in the context of large-scale
digital surveys the task of visually checking many targets against
numerous published finding charts is discouragingly onerous, a
problem common to all the early major surveys.  The situation has
gradually improved in recent years, particularly as Brian Skiff (Lowell
Observatory) has obtained modern identifications, and hence precise
astro\-metry, for the major photo\-graphic objective-prism surveys of
LMC targets, as one aspect of his extensive and continuing on-line
Catalogue of Stellar Spectral Classifications, commonly cited as
`B/mk' \citep{Skiff13}. His efforts
have provided precise positions for the \citet{Henize56} listing, among
others, while \citet{Howarth12b} has given identifications and
astro\-metry for the BE74 catalogue.

The last (and largest) of the major emission-line catalogues in need
of systematic, precise positions is the Armagh survey.  This was
published in three tranches; the first contains objects with no
detectable continuum, interpreted as emission nebulae
\citep{Lindsay63a}.  The second and third lists are of ``objects of
stellar appearance'' (\citealt{Lindsay63b}, L63; \citealt{Andrews64},
AL); it is these stellar sources that are the principal subject of
this Note.

\section{Methods}

\citet{Lindsay63b} and \citet{Andrews64} initially published only
approximate co-ordinates, but, after Lindsay's death, finder charts
were published at Armagh Observatory by McFarland
(\citealt{McFarland75}, hereinafter MLA), in the form of a folder of
loose-leaf reproductions of blue (103a-0) plates at a scale of
$\sim{16}\arcsec$/mm.

Using these charts, counterparts of the Armagh emission-line stars
have been visually identified on red images of the second Digitized
Sky Survey (DSS2), using the CDS's {\sc Aladin} tool;  
where necessary, identifications were refined as described in
Section~\ref{sec:xid}.
The corresponding
positions were recorded inter\-actively, and transferred to a data file
by copy-and-paste.  Correlating the results against UCAC4
\citep{Zacharias12} gave a positive match with a single target within
5$\arcsec$ in most cases, with a sub-arcsecond systematic offset.
After correcting the inter\-actively recorded measurements for this
offset, a second pass was made against UCAC4 with a 2$\arcsec$-radius
window to obtain final positions.  Positional differences between
corrected inter\-active measurements and UCAC4 are less than one
arc-second in the great majority of cases.

The results are listed in Table~A1 (on-line); cross-matches against the UCAC2
and 2MASS catalogues are included for the convenience of those who
prefer these better-established sources (including consistency with
B/mk), although the positional precision is now such that
cross-matching against any catalogues of interest should be
straightforward.

\section{Cross-identification metacatalogue}
\label{sec:xid}

The co-ordinates reported here are principally intended to establish
precise positions for the objects marked by MLA on their finder
charts.  However, in a number of cases the charts present ambiguities
or other uncertainties in identification,\footnote{One
contributing 
factor is
  that the scale and
  image quality of the MLA charts mean that it is not feasible to
  resolve stars separated by less than $\sim{10}\arcsec$, at best
  (cp.\ DSS2 imaging, which allows multiple objects to be identified
  at separations down to $\lesssim{3}\arcsec$).} and in an effort to
alleviate these uncertainties and clarify correct identifications, the
new Armagh results have been correlated with the \citet{Henize56} and
\citet{Bohannan74} lists, using the Skiff and Howarth
positions, but referring back to the original finder charts
as necessary.

The BAT99 catalogue of 134 LMC Wolf-Rayet stars \citep{BAT99} and the
recent \citet{Reid12} catalogue of H$\alpha$ emission-line stars (576
RPs stars) were also used in this effort, with matching based on
positional coincidence alone; for both lists, good astrometry is available from
the original sources.\footnote{For convenience, the BAT99 source positions
  were actually taken from \citet{Bonanos09}, which has reliable, precise
  astrometry for all these targets (though their compilation of LMC
  spectral types is much less complete than B/mk).}  For completeness,
the BAT99 listing was augmented by the five subsequent LMC WR
discoveries announced up to the end of 2012 \citep{Evans11, Gvaramadze12, 
Howarth12a, Neugent12},\footnote{In addition,
  \citet{Massey00} newly identify Sk $-69^\circ$ 194 (L63~289) as
  a WR star.} although no effort
has been made to include individual stars from other sources (not
least because of the ambiguity of what properly constitutes an
`emission-line' star in the present context).

The resulting metacatalogue amalgamates all the major
spectroscopically-based surveys dedicated to finding LMC emission-line
candidates, and yields 1675 stars that (probably) represent separate
objects.  The metacatalogue is given in Table~A2, with comments in
Table~A3 (both on-line), while Table~\ref{tab:xmatch} summarizes the
contributions from the various primary sources.  Since many targets
are known under several names, a unique running `LMCe' number is assigned
in Table~A2, for internal cross-referencing to other Tables (and not
to promulgate yet another label for objects which already all have at
least one unique identifier).

\begin{figure}
\centering
\includegraphics[scale=0.43,angle=0]{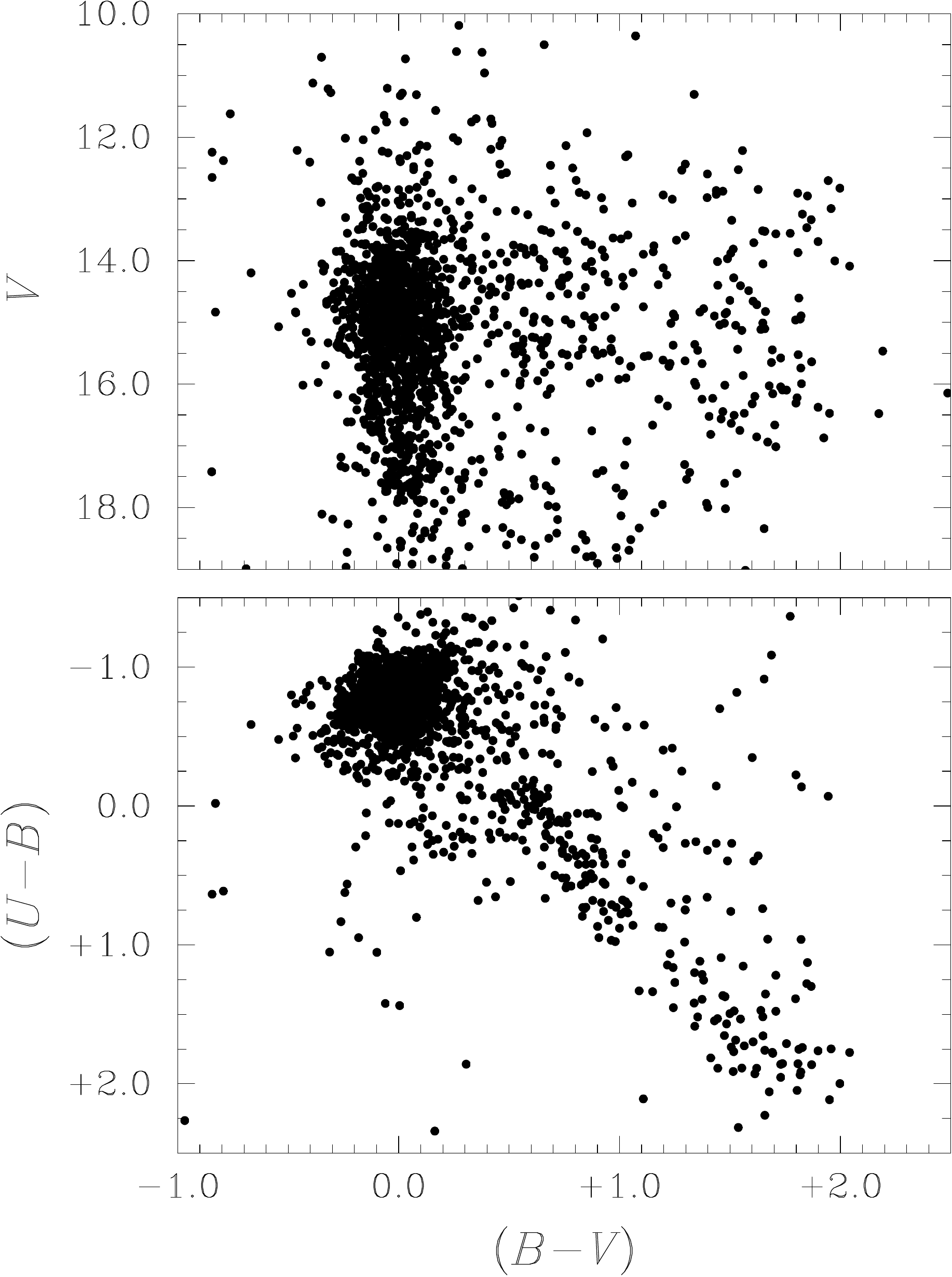}
\caption{Colour--magnitude and colour--colour diagrams for entries in
  the consolidated metacatalogue (MCPS photo\-metry;
  \citealt{Zaritsky04}).}
\label{fig1}
\end{figure}

Cross-matches of the metacatalogue entries have been made with the
Reid \& Parker planetary-nebula survey \citep{Reid06}, and the
\citet{Sabogal05} catalogue of Be-star candidates (proposed on the
grounds of characteristic photo\-metric variability); this helped to
resolve identification ambiguities in a few cases (particularly in
crowded fields).  For completeness, matches have also been made with
the Henry Draper Extension (HDE;
\citealt{Cannon36}), the Sanduleak catalogue (Sk;
\citealt{Sanduleak70}), the Radcliffe luminous-star catalogue (RMC;
\citealt{Feast60}), and the VLT-Flames Tarantula Survey (VFTS;
\citealt{Evans11}).   Results are included in Table~A2.

\citet{Duflot10} has compiled a related metacatalogue of
  LMC stars, of all types.  His effort differs from that reported here
  both in its more ambitious scale (4011 entries), and in that he
  appears to rely on the co-ordinates published in primary sources as
  the main criterion for identifying catalogue matches (leaving the
  question of the actual stellar identifications unresolved in many
  cases, although Skiff has since provided modern identifications, and
  precise astrometry, for most of the targets).  The current
  metacatalogue has been compared with Duflot's; results are included
  in Table~A2, and discrepancies discussed in Table~A3.  In general,
  the agreement in matchings of primary catalogues is remarkably good;
  in only three cases does Duflot appear to assign physically different stars to
  a single identifier (see Table~A3 entries for LMCe 951, 1282, and
  1285), while overlooking 10 matches identified here, and omitting
  two stars (AL-29 and BE74-527).  Reassuringly (and surprisingly!),
  no revisions to the composition of the emission-line metacatalogue
  were required as a result of the comparison with \citet{Duflot10},
  encouraging a view that probably no more than a handful of
  straighforwardly identifiable errors remain.

\subsection{Overview of metacatalogue properties}

Figure~\ref{fig1} shows colour--magnitude and colour--colour diagrams
for the metacatalogue stars.  The photometry is taken from
\citet{Zaritsky04}, to whom reference should be made for data-quality
caveats. As expected, the samples are evidently dominated by luminous,
blue stars, although a number of redder objects occur.  While many
of these redder objects may have have intrinsic H$\alpha$ emission,
the objective-prism surveys may also include
M-type stars in which the TiO bandhead at 654nm `mocks'
H$\alpha$ emission (\citealt{MacConnell83};  Skiff, personal communication).

\begin{figure}
\centering
\includegraphics[scale=0.43,angle=0]{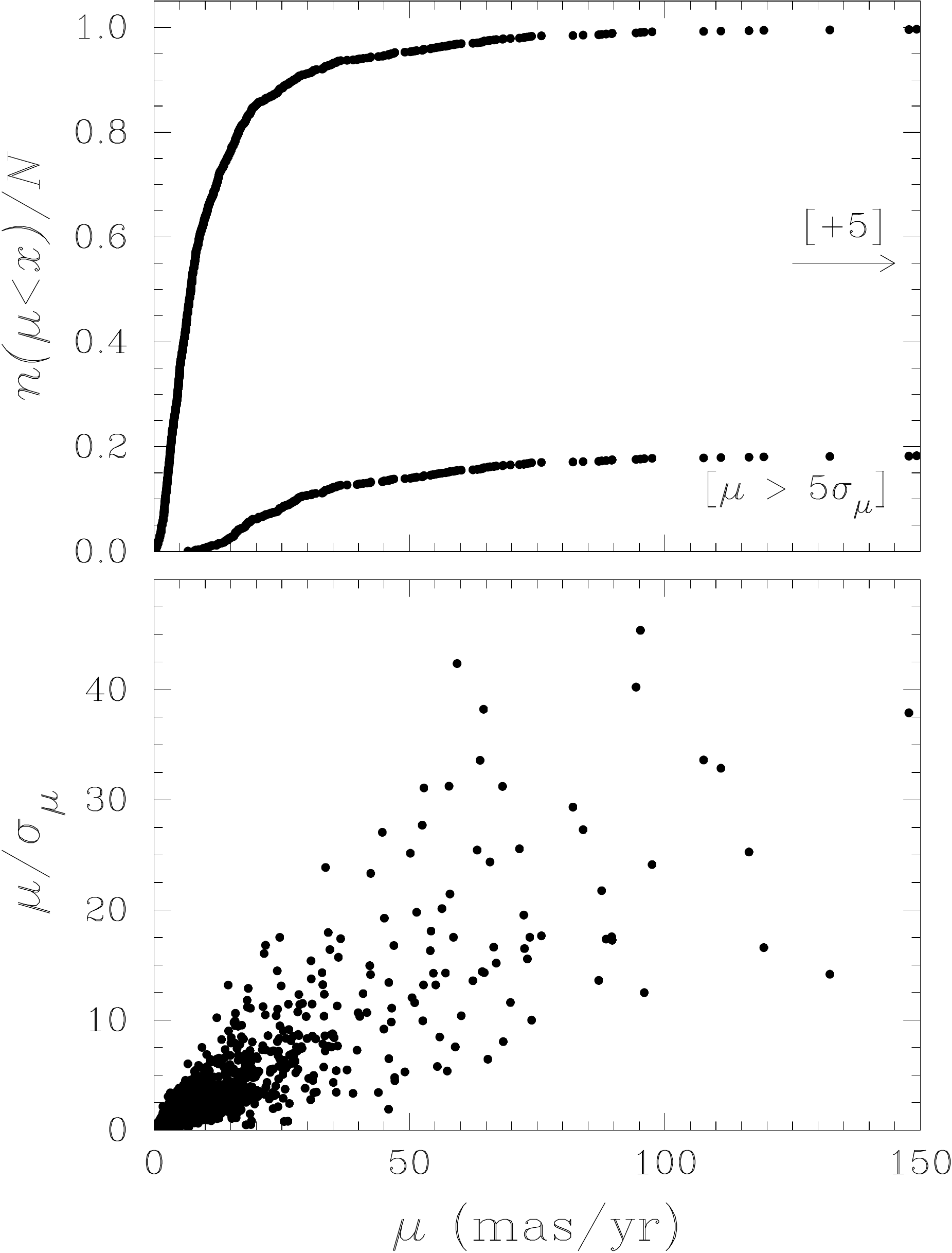}
\caption{Upper panel: cumulative distribution function of
UCAC4 proper motions for 1417 stars in the consolidated metacatalogue, 
including the subset of 264 objects where the proper motion is more than
$5\times$ its error. Lower panel:  detection significance.}
\label{fig2}
\end{figure}

UCAC4 proper motions are available for 1417 of the entries in
Table~A2, with a median error of $3.6 \mbox{ mas yr}^{-1}$. LMC
members are not expected to show detectable proper motions at this level of
accuracy; in practice, 264 stars have proper motions that are
statistically formally significant at the 5-$\sigma$ level or above,
typically corresponding to $\mu \gtrsim 17 \mbox{ mas yr}^{-1}$
(Fig.~\ref{fig2}).   Many of these will be foreground objects, although
some spuriously high proper motions must arise through errors.   UCAC4
proper-motion information is included in Table~A2, for guidance.

\section{Discussion}

The newly available redundancy among primary catalogues was often
crucial in determining the correct identification of an emission-line
source (or, at the least, in clarifying where ambiguities remain), as
elaborated on a case by case basis in Table~A3.  Consequently,
27 (4.3\%) of the BE74 identifications adopted by
\citet{Howarth12b} are revised, as summarized in Table~A4 (on-line).
Two of these revisions arise from gross errors (wrong star originally
recorded), 13 from new or revised resolutions of ambiguities in the
BE74 charts, and 12 from newly revealed probable errors by BE74.

The relatively bright Wolf-Rayet star \mbox{BAT99~131} offers an
interesting, and cautionary, illustration. \citet{Westerlund64}
discovered it under the name \mbox{WS 51}, and identified it
(proleptically) with UCAC4~115-010182.  \citet{Bohannan74}
independently recovered it under the guise of \mbox{BE74~151}, marking
the same star on their charts as did \citeauthor{Westerlund64}.
Meanwhile, \citet{Sanduleak70} had marked his \mbox{Sk~$-67^\circ$~259} as a
star of similar brightness lying 13$\arcsec$ to the SW (UCAC4
115-010180), identifying it with the WR; but \citet{Fehrenbach76}
later asserted that Sanduleak's identification was in error.
\citet{Breysacher81} obtained new slit spectroscopy of the WR, as Brey~98, and
implicitly confirmed the identification with the NE star (repeating
Fehrenbach et al.'s assertion, although identifying Brey~98 with, inter alia,
both \mbox{Sk~$-67^\circ$~259} \emph{and} WS~51).

However, \citet{McFarland75} identify AL~412 with the SW star (i.e.,
with \mbox{Sk~$-67^\circ$~259}), and \citet{BAT99} give a finder
chart, and precise astrometry, also identifying this as the WR,
without further comment.  \citet{Foellmi03b} and \citet{Crowther06}
observed the star; the former reproduce the BAT99 co-ordinates (again
without comment), and
Crowther (personal communication) confirms, on the basis of his finder
charts and telescope pointing, that the SW component is a WR star.
Either both stars are emission-line objects (both are blue in $B-V$),
but each investigation found only one; or, more plausibly, both
\citet{Westerlund64} and \citet{Bohannan74} are in error, as is the
`correction' in \citet{Fehrenbach76}.

No doubt similar cases remain to be resolved (several potential
instances are discussed in Table~A3) but, as in this example, new
spectroscopy is likely to be the only secure method of addressing
these issues in general.

\acknowledgements{The HDE, RMC, LH$\alpha$~120-S, and Sk co-ordinates
used here are all Brian Skiff's work.
I thank him for helpful correspondence, and
  not least for bringing the MLA charts to my attention.   John
  McFarland \& Mark Bailey, of Armagh Observatory, kindly provided
  both a set of those charts, and a pdf of a high-quality scan
  (subsequently made more generally available by them, through the
  ADS).  The referee, Fran\c{c}ois Ochsenbein, suggested the
  comparison with the Duflot compilation.
The execution of this work relied on resources and tools
  provided by the Centre de Donn\'ees astronomiques de Strasbourg.

\smallskip\noindent
Until publication in CDS, on-line tables will be available at 
\href{ftp://ftp.star.ucl.ac.uk/idh/LMCe}{\tt {ftp://ftp.star.ucl.ac.uk/idh/LMCe} }
}
\bibliographystyle{aa}

\bibliography{IDH}

\end{document}